\documentclass[seceq]{ptptex}

\usepackage{graphicx}

\title{
Molecular Transport through a Bottleneck Driven by External Force
}

\author{
Chihiro \textsc{Nakajima}
and Hisao \textsc{Hayakawa}%
}

\inst{
Yukawa Institute for Theoretical Physics, \\
Kyoto University, Kyoto 606-8502, Japan 
}

\recdate{July 27, 2009; Revised September 16, 2009}

\abst{
The transport phenomena of Lennard-Jones molecules through
a structural bottleneck  
driven by an external force
are investigated by molecular dynamics simulations. 
We observe 
two distinct molecular flow regimes distinguished by 
a critical external force $F_{c}$
and find 
scaling behaviors between external forces and flow rates.
Below the threshold $F_{c}$,
molecules are essentially stuck in the bottleneck
due to the attractive interaction between the molecules,
while above $F_{c}$, molecules can smoothly move in the pipe.
A critical flow rate $q_{c}$ corresponding to $F_{c}$
satisfies a simple relationship with angles,
and the value of $q_{c}$ can be estimated from a simple argument.
We further clarify the role of the temperature dependence  in the molecular flows
through the bottleneck.
}

\begin{document}
\maketitle

\section{Introduction}
Transport phenomena through a structural bottleneck have been widely observed 
and have attracted attention owing to the development of nanoscience.
Many interesting bottleneck features have been observed and studied in 
granular flows \cite{Pennec,Longhi},
traffic flows \cite{Helbing_RMP,Nagatani_Rep,yamamoto},
evacuation processes  \cite{Kretz, Isobe, Nagai},
water flow through porous grains \cite{Carminati},
microfluidics \cite{Burriel},
Laval nozzles \cite{laval}, and
the Internet \cite{Sameet}.
Recent technical advances  have enabled us to study 
the nanoscale transport system such as
ion transport through protein channels in a lipid membrane 
\cite{Doyle, Bonthuis, Zhang},
artificial nanopores \cite{Wallacher},
carbon nanotubes \cite{Berezhkovskii},
Knudsen gases \cite{knudsen,nishino},
and 
domain wall motion in a wire with a notch \cite{koyama,himeno,tanigawa,Konig}.
Thus, a comprehensive understanding of transport phenomena
through a bottleneck is essential.

Unlike bulk systems, the bottleneck alters the transport properties.
Jamming phenomena relating 
the formation of a permanent arch at the bottleneck 
\cite{To}
and intermittent outflows in a funnel with various angles \cite{horluck}
have been observed 
and intensively studied
\cite{Helbing}.
In electrical transport through quantum wires, the effect of the 
geometry of the wire 
on the transmission coefficient and conductance 
has been an important issue \cite{vargiamidis,midgley}.
Recently, 
experiment and simulation studies have been performed
to understand the structure and dynamics of water in a confined carbon nanotube,
where it is discussed
that  confinements induce the 
phase transition not observed in a bulk system \cite{Koga,Hummer}.
In the transport phenomena through the bottleneck,
these collimator effects such as the curvatures and  cross sections of the passage
play a crucial role in the transport phenomena.
In such situations, 
a useful approach to treat transport phenomena through a bottleneck
is computer simulations.
Various simulation techniques, such as 
driven lattice gases \cite{Pierobon, philip}, 
cellular automata \cite{Popkiv}, and 
molecular dynamics (MD)\cite{Jepps}
have been designed and applied to probe these transport mechanisms.
Numerically,  
the existence of scaling behavior of escape dynamics through  
a bottleneck \cite{Tajima, Nagatani}
and 
the power spectra of the flow in an hourglass \cite{veje}
have been investigated.

In this paper, we address the problem of the steady particle flow in a pinched
pipe under an external force. 
We restrict our attention to a highly simplified two-dimensional model, considering geometric effects
in the hope to gain some generic physical insight.
We intend to evaluate the importance of 
the curvature of the pipe,
particle area fractions, and external forces.
This goal may be achieved through detailed simulations of the motion of particles.
For these reasons, we perform MD simulations to investigate molecular flows
in the pipe with various angles.

The present article is organized as follows.
In $\S$\ref{models}, we discuss the  model and MD simulation techniques.
In $\S$\ref{steady}, the results of the MD simulation and  data analyses 
are presented. 
Special emphasis is placed on the relationship between 
stationary molecular flows  and external forces at various angles
and scaling behavior of these relations.
Then, we demonstrate that flow rates against angles 
give a simple relation and estimate the value of the critical flow rate.
We  investigate the fluctuation of the flow and the fundamental diagram for molecular flows in analogy to a traffic model.
We further investigate the temperature dependence of the molecular flow
and how flow rates depend on the ratio between  molecular size and  pipe
width.
For comparison, we observe a molecular flow
that has only repulsive interactions.
We conclude and discuss our results in $\S$\ref{con}.

\section{\label{models}Model}
\begin{figure}
\begin{center}
\includegraphics[width=11cm]{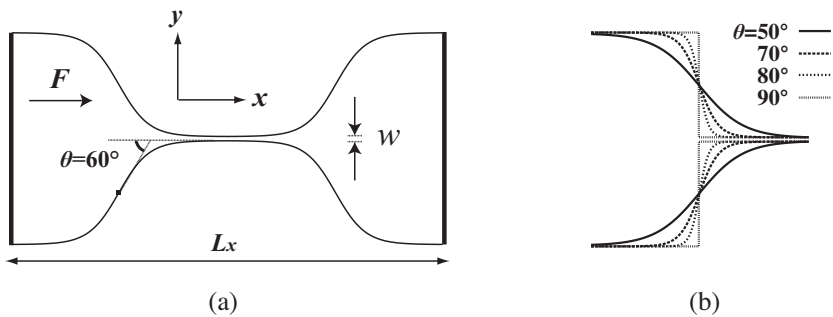}
\caption{\label{figure} 
(a) Schematic of a two-dimensional symmetric pipe with angle $\theta$, 
which is characterized by the slope of $y=36\tanh(x\tan\theta/36)$.
The $x$-directional pipe length is $Lx=300 \sigma$ and the narrowest pipe width
is $w$. 
The wall is constructed of particles that are attached to other particles by a linear spring,
and the $y$-directional boundary is the thermal wall with temperature $T=0.1\varepsilon$.
(b)
Schematic of left region of the pipe with angle $\theta=50^{\circ}$, $70^{\circ}$, $80^{\circ}$, $90^{\circ}$.
}
\end{center}
\end{figure}

Let us consider a two-dimensional pinched pipe shown in Fig. \ref{figure}(a).
The pipe has symmetry configuration, like an hour glass,
and is represented by the hyperbolic tangent $y= 36\tanh(x\tan\theta^{\circ}/36)$
with the angle $\theta$ defined by the slope of hyperbolic tangent.

The  molecules are interacting via Lennard-Jones (LJ) potential:
\begin{equation}
\phi_{\text{LJ}}(r_{ij}) = 4\varepsilon \left\{\left(\frac{\sigma}{r_{ij}}\right)^{12}
-\left(\frac{\sigma}{r_{ij}}\right)^{6}\right\},
\end{equation}
where $\varepsilon$ is the energy scale of the interaction, 
$\sigma$ is the diameter of each molecule,
and $r_{ij}$ is the separation between particles $i$ and $j$.
The pair interactions are truncated at $r_{ij}=3\sigma$ 
for numerical convergences.
The linear $x$-directional pipe length is $L_{x}=300\sigma$.
In most of our simulations,
the narrowest pipe width $w$ is 3$\sigma$,
but we will later investigate the width dependences on the flow rate.
The wall is composed of molecules that are attached to other molecules 
by linear springs with spring constant
$\kappa=20000\varepsilon/\sigma^{2}$, 
but the molecules at both edges are fixed.
The wall molecules are interacting with driven molecules via the same LJ potential.
In the presence of the $x$-directional external force $F$,
the current of the molecules will be maintained, and 
hence, the system will always remain in a nonequilibrium stationary state.

In the horizontal boundaries, 
we adopt the  periodic boundary conditions for the positions of molecules
and the thermal walls for the velocity of molecules,
which give the random velocity whose probability distribution function is
given by
\begin{equation}
  \label{eq:netsu}
  f(\boldsymbol{v})=\frac{1}{\sqrt{2\pi}}\left(\frac{m}{k_{B}T}\right)^{3/2}|v_{x}|e^{-\frac{mv^2}{2k_{B}T}}
\end{equation}
with temperature $T=0.1\varepsilon$. 
Owing to the presence of  the $x$-directional external forces, 
molecules keep flowing from left to right 
through the bottleneck.

For comparisons of LJ molecular flows,
we adopt the purely repulsive Weeks-Chandler-Andersen (WCA) potential 
given by 
\begin{equation}
\phi_{\text{WCA}}(r_{ij})=
\begin{cases}
4\varepsilon
\left\{\left(\frac{\sigma}{r_{ij}}\right)^{12}-\left(\frac{\sigma}{r_{ij}}\right)^{6}
+ \frac{1}{4}\right\}&,(x\le2^{1/6}\sigma)\\
0&,\text{otherwise}
\end{cases}
\label{WCA}\\
\end{equation}
which we call WCA for later discussion.

MD simulation is carried out for the dynamics of the molecules 
under the different geometrical pipes.
The equation of motion is solved by using the velocity-Verlet method 
with a time step $\Delta t=0.001\tau$, where $\tau=\sigma(m/\varepsilon)^{1/2}$.
The mass of the driven molecule is  $m$ and that of the wall is  $10m$ to avoid
the change in the pipe configuration due to the molecular flow through the pipe.
The initial positions of molecules are arranged not to contact each other,
and the initial velocities of the molecules are given by the Maxwell distribution
with temperature $T=0.1\varepsilon$.
In our reduced units, the unit length $\sigma$, 
unit energy $\varepsilon$,
molecular mass $m$, and Boltzmann constant $k_{B}$ 
are set equal to unity.
As shown in Fig. \ref{figure}(b),
we vary the pipe angle $\theta=50^{\circ}$, 55$^{\circ}$, 60$^{\circ}$, 65$^{\circ}$, 70$^{\circ}$, 75$^{\circ}$, 80$^{\circ}$, 85$^{\circ}$, 90$^{\circ}$.
As the angles become large,
each pipe represents a narrow straight passage with 
a narrowed entrance and constant pipe width,
whereas 
the pipe with a small angle exhibits a pinched configuration.
The pipe configuration less than angle $\theta=50^{\circ}$ 
is not used since the $y$-directional pipe length
becomes markedly short in our model.

To investigate molecular flows in the  pinched pipe,
we mainly measure the flow rate $q$ in the pipe with various angles, 
which is the numbers of crossing-driven molecules
during the time at the middle cross section.
In the steady flow, the flow rate $q$ should
be a constant in any cross section owing to the continuity equation in the stationary state. 
After $t=1.5\times10^{4} \tau$, 
we collect the statistical quantities
as the system reaches a steady state.

\section{\label{steady}Steady molecular flow}
\subsection{Steady molecular flow at various angles}
\begin{figure}
\centerline{
\includegraphics[width=9cm]{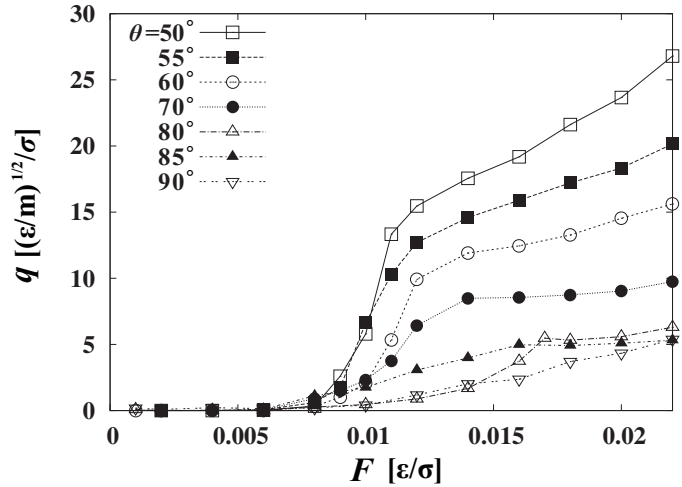}}
\caption{\label{qF} 
Plots of the flow rates $q$ against the various external forces $F$ in the
pipe with angle $\theta=50^{\circ}$, 55$^{\circ}$, 60$^{\circ}$, 70$^{\circ}$, 80$^{\circ}$, 85$^{\circ}$, 90$^{\circ}$.} 
\end{figure}

\begin{figure}
\centerline{
\includegraphics[width=9cm]{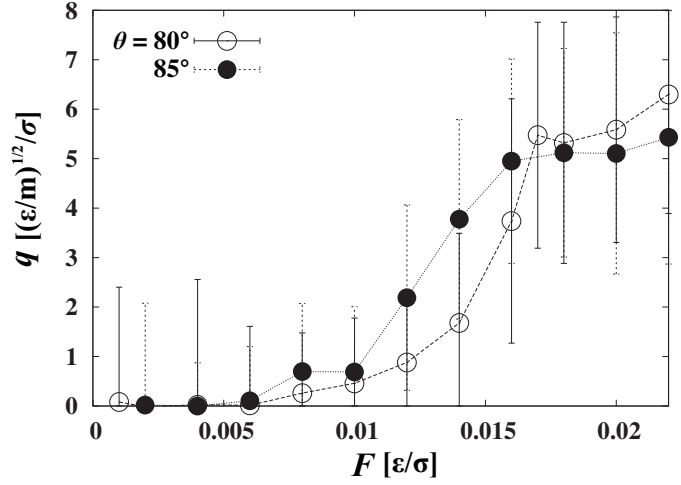}}
\caption{\label{qF_yerro} 
Plots show the flow rates $q$ against the external forces $F$ at the  angle $\theta=80^{\circ}$
and 85$^{\circ}$ with variance.} 
\end{figure}

\begin{figure}
\centerline{
\includegraphics[width=8cm]{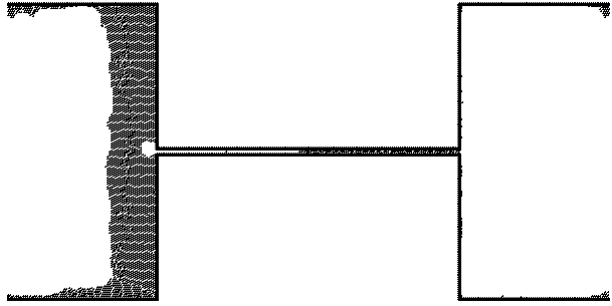}}
\caption{\label{90} A schematic pipe graph at the angle $\theta=90^{\circ}$ and the
  external force $F=0.010\varepsilon/\sigma$.
The entrance of the pipe exhibits arch formation.}
\end{figure}

We study steady molecular flows through the pipes with various angles.
The narrowest width of  each pipe is three times  the molecule diameter.
The driven molecule number in the pipe is $N=3510$.
Figure \ref{qF} plots the flow rate $q$
against the external force $F$ under various angles.
In the case of small external forces,
driven molecules jam in the pipe and stick on the wall
since the molecular interactions are sufficiently strong compared with external forces,
while as the external forces increase, 
the nonlinear flow growth is observed 
and jamming is insignificant owing to the dominance of the external forces.
These transport phenomena are clearly observed in the pipes with angle 
below $\theta=80^{\circ}$.
On the other hand, as Fig. \ref{qF_yerro} shows, 
the distinct relation between the flow rate and external force
is difficult to observe  at the angle above $\theta=80^{\circ}$
owing to the small amount of molecular flow caused by the markedly narrow 
pipe entrance and the passage at these angles.
At angle $\theta=90^{\circ}$, 
the molecular arch
is observed in the low external force. 
Figure \ref{90} exhibits the existence of a  molecular arch in the entrance of the pipe
with angle $\theta=90^{\circ}$ under the external force $F=0.010\varepsilon/\sigma$.
This phenomenon could be caused by 
the straight corner at the pipe entrance.

\begin{figure}
\centerline{
\includegraphics[width=9cm]{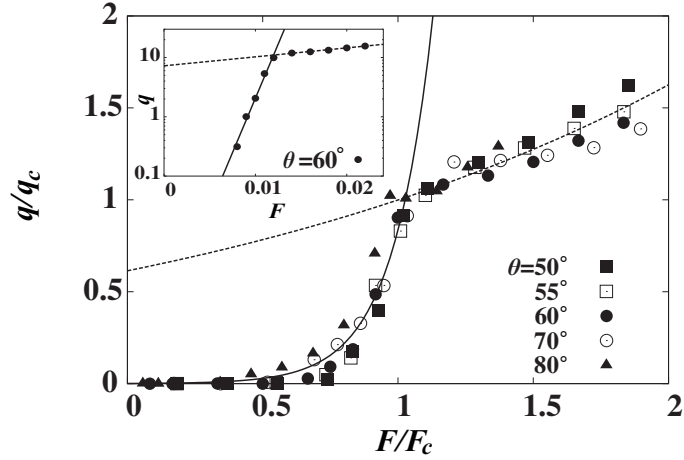}}
\caption{
Plot of the scaled flow $q/q_{c}$ against the scaled external force $F/F_{c}$ at the 
angle $\theta=50^{\circ}$, 55$^{\circ}$, 60$^{\circ}$, 70$^{\circ}$, 80$^{\circ}$.
The solid  line is fitting curve $q/q_{c}=\exp(a F/F_{c}-b)$
with $a,b=(8.45, 8.48)$ for molecular flow below $F_{c}$ and 
the dotted line is $a,b=(0.445, 0.431)$ for molecular flow above $F_{c}$.
\label{scale} The inset shows a semilog plot of the flow rate and the external force in
  the pipe with angle $\theta=60^{\circ}$. 
We define the crossover of two flow lines as $F_{c}$ and $q_c$.
}
\end{figure}

To investigate the molecular flow properties further,
we perform the semilog plot of the flow rate, $\log q$, against the external force $F$.
The inset of Fig. \ref{scale} indicates the existence of 
two distinct molecular flow regimes 
in the pipe with angle $\theta=60^{\circ}$.
The flow rate and external force at the crossover of these regimes, 
$q_c$ and $F_c$,
are considered to correspond to the balance of external forces and molecular interactions.
These transport phenomena are observed at the angle below $\theta=80^{\circ}$ 
and $q_{c}$  and $F_{c}$ are determined at each angle.
At the angle above $\theta=80^{\circ}$, 
it is difficult to distinguish characteristic crossover
from the relation between flow rates and external forces.  
By using $q_c$ and $F_c$ in each pipe,
the scaled flow $q/q_{c}$ against the scaled external force $F/F_{c}$ 
can be plotted in Fig. \ref{scale}. 
The solid and dotted lines are fitted by curves $q/q_{c}=\exp(a F/F_{c}-b)$
with $a,b=(8.45, 8.48)$ for molecular flow below $F_{c}$ and 
$(0.445, 0.431)$ for molecular flow above $F_{c}$, respectively.
Thus, it is shown that
flow rates as  a function of external forces 
are given by the two distinct simple exponential forms 
and are independent of the pipe geometry
in these regimes.

\begin{figure}
\centerline{
\includegraphics[width=9cm]{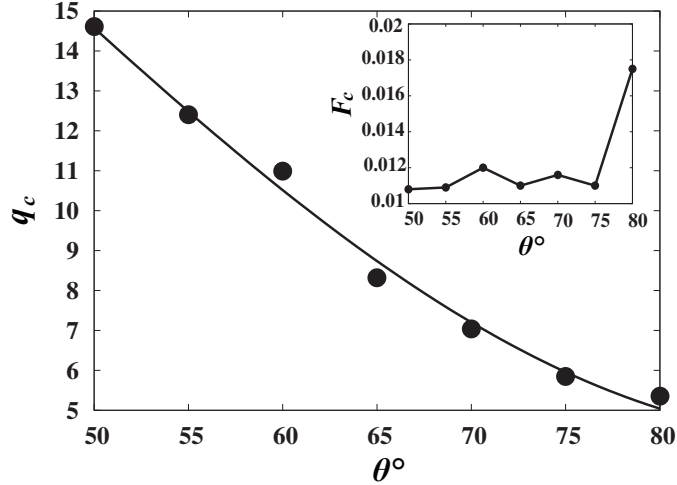}}
\caption{\label{qtheta} Critical flow rate $q_{c}$ as a function of the pipe angle
$\theta$. The solid line is the fitted curve
$\alpha \cos^{2} \theta +q_{0}$ with $\alpha=25.9(\varepsilon/m)^{1/2}/\sigma$ and 
$q_{0}=4.21(\varepsilon/m)^{1/2}/\sigma$.
The unit of the vertical axis is $(\varepsilon/m)^{1/2}/\sigma$.
The inset shows the plot of the critical external force $F_{c}$ [$\varepsilon/\sigma$] against the angle $\theta$.}
\end{figure}

The angle dependences on $q_c$  can be simply studied.
Figure \ref{qtheta} is the plot of $q_c$ as a function of the angle.
At a large angle,  
pipe configurations represent the reduced passage
so that 
it is plausible for the molecular flow to decrease against the angles.
We are aiming at a rough theoretical estimation of 
variations of $q_c$ against the angles.
In the presence of $F_c$, the system is considered to be  balanced 
between the external force and the molecular interaction.
We assume that the practical external force along the pipe with angle $\theta$, $F_{c}\cos\theta$,
drives molecules in the pipe.
This practical external force is set off in 
the $y$-direction owing to the symmetric pipe configuration;
thus, the $x$-component of the practical external force $\tilde{F}\simeq
F_{c}\cos^{2}\theta$
could be dominant for molecular flows.
Since we consider the steady state,
the particle density in the filled pipe region $\bar{\rho}$ 
and the terminal velocity $v \simeq \mu \tilde{F}$ with the mobility $\mu$
are considered to be  constant.
Thus, the flow rate at the balanced point is approximated by
\begin{equation}
 q_{c}\simeq \alpha \cos^{2}\theta + q_{0}, 
\label{fitting}
\end{equation}
with fitting parameters $\alpha$ and $q_{0}$, 
where $q_{0}$ is the extrapolation of the critical flow rate at $\theta=90^{\circ}$.
In Fig. \ref{qtheta}, 
the solid line represents \eqref{fitting} with
$\alpha=25.9(\varepsilon/m)^{1/2}/\sigma$ and $q_{0}=4.21(\varepsilon/m)^{1/2}/\sigma$ and is well fitted at the angle below
$\theta=80^{\circ}$.
On the other hand, 
the relation $F_{c}$ and the angle
are almost independent of the angle under $\theta \le 75^{\circ}$ in the inset of Fig. \ref{qtheta}.
At the angle above $\theta=80^{\circ}$,
the value of $F_{c}$ that corresponds to the value of $q_{c}$
cannot be determined since the crossover is not clearly observed.   

We further estimate the value of $q_{c}$, which corresponds to $F_{c}$.
The critical flow rate $q_{c}$ is given by
\begin{equation}
q_{c}=nv_{c}S,
\end{equation}
where $n=4\phi/(\pi \sigma^{2})$ is the number density with the area fraction $\phi$, 
$v_{c}$ is the critical velocity, and $S$ is the cross section.
 The critical velocity is represented by the escape velocity from 
 the bottom of LJ potential as 
$v_{c}=\sqrt{2|\phi_{LJ}(a)|/m}$
with the distance at the minimum potential $a=2^{1/6}\sigma$.
Thus, the critical flow rate can be estimated by 
 \begin{equation}
q_{c}=\frac{12\sqrt{2}}{\pi }\frac{1}{\sqrt{m\sigma}}\phi.
\end{equation}
The area fraction at the region where the particles accumulate
is $\phi \simeq 0.7$ in our model, 
so that we can estimate the value of the critical flow rate as $q_{c}\simeq 3.8\sqrt{\varepsilon/m}/\sigma$. 
This value is close to the critical flow rate $q_0$ at the angle $\theta=90^{\circ}$.
Thus, these theoretical estimations are plausible for the explanation of our molecular transport.

\subsection{Fluctuation of the flow}
\begin{figure}      
\centerline{        
\includegraphics[width=9cm]{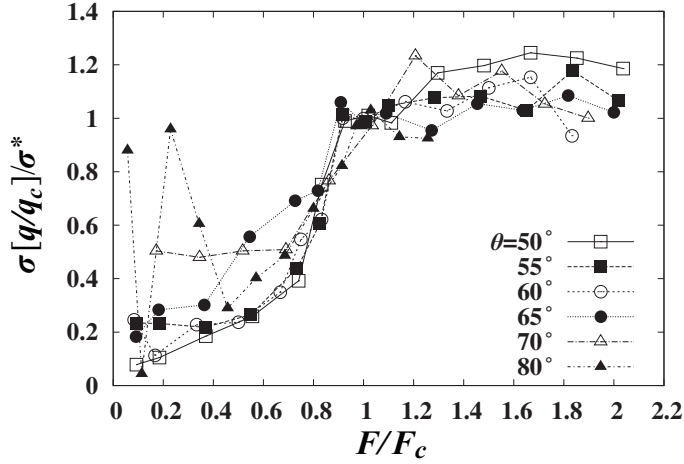}}
\caption{Standard deviation of the scaled flow rate $\sigma[q/q_{c}]/\sigma^{\ast}$ against the scaled external force $F/F_{c}$ at the angle $\theta=50^{\circ}$, 55$^{\circ}$, 60$^{\circ}$, 65$^{\circ}$, 70$^{\circ}$, 80$^{\circ}$. The normalized standard deviation $\sigma^{\ast}$ is the
linear interpolation of the standard deviations $\sigma[q_{1}/q_{c}]$ and $\sigma[q_{2}/q_{c}]$, 
where $q_{1}$ and $q_{2}$ are the closest points from $q_{c}$ ($q_{1}\le q_{c}\le q_{2}$)
in each angle.}
\label{3}
\end{figure}

We next investigate how the fluctuation of the flow depends on the external force.
To clarify the characteristics of  the fluctuation,  we introduce the normalized standard deviation $\sigma^{\ast}$
obtained by the  linear interpolation of the standard deviations $\sigma[q_{1}/q_{c}]$ and $\sigma[q_{2}/q_{c}]$, 
where $q_{1}$ and $q_{2}$ are the closest points from $q_{c}$ ($q_{1}\le q_{c}\le q_{2}$).
Figure \ref{3} shows the standard deviation of the scaled flow rate 
$\sigma[q/q_{c}]/\sigma^{\ast}$ against the scaled external force $F/F_{c}$ 
at the angle $\theta=50^{\circ}$, 55$^{\circ}$, 60$^{\circ}$, 65$^{\circ}$, 70$^{\circ}$, 80$^{\circ}$.
As Fig. \ref{3} shows, the scaled standard deviation $\sigma[q/q_{c}]/\sigma^{\ast}$ 
increases gradually below the external force $F<F_{c}$,
where the particles are essentially stuck in the bottleneck 
and the fluctuations are suppressed.
For the large external forces $F>F_{c}$ 
the scaled standard deviation is almost constant in any pipe angle and 
the fluctuation of the flow is insensitive to  the external force.
These fluctuation phenomena almost correspond to Fig. \ref{scale}
(the relation between the scaled flow rate and the scaled external force).
It is interesting that the fluctuation of the flow does not 
have any specific peak around $F_{c}$, and the critical fluctuation continues
above $F>F_{c}$, while we do not make clear its reason.

\begin{figure}      
\centerline{        
\includegraphics[width=9cm]{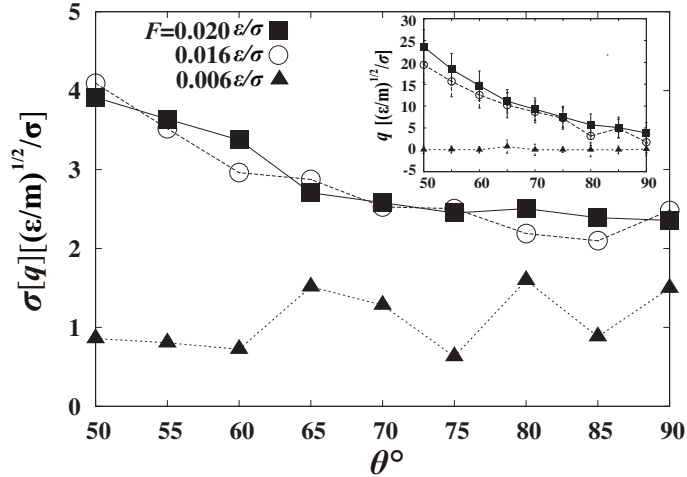}}
\caption{Standard deviation of the flow rate $\sigma[q]$ against the angle $\theta$ 
in the external force 
  $F=0.020, 0.016, 0.006 \varepsilon/\sigma$. The inset shows  the corresponding flow rate 
  $q$ against the angle $\theta$.}
\label{kakudo_eroo}
\end{figure}

Furthermore, we investigate how the fluctuation depends on the angle.
Figure. \ref{kakudo_eroo} shows the standard deviation of the flow $\sigma[q]$ 
against the angle $\theta$ at the constant external forces $F = 0.020, 0.016, 0.004
 \varepsilon/\sigma$,
and the inset of Fig. \ref{kakudo_eroo} shows
the corresponding flow rate $q$ against the angle $\theta$.
In each external force, the standard deviation of the flow rate 
is substantially constant and does not exhibit any fluctuation peak against the angle.
Although we have investigated  the relation between the fluctuation and the angle,
the angle dependence of the fluctuation of the flow is not observed in our model.

\begin{figure}      
\centerline{        
\includegraphics[width=9cm]{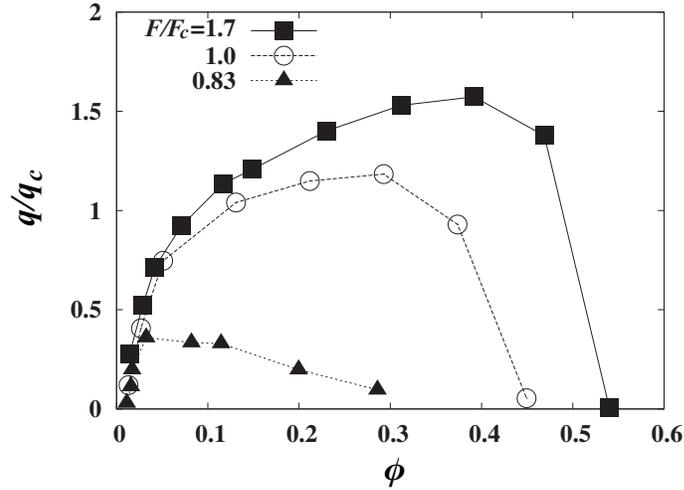}}
\caption{\label{funda} Scaled flow rate $q/q_{c}$ as a function of the  area
fraction $\phi$$[1/\sigma^2]$ under scaled constant external forces
$F/F_{c}=0.83, 1.0, 1.7$ with $F_{c}=0.012 \varepsilon/\sigma$.}
\end{figure}

\subsection{Fundamental diagram}
For traffic flows,
the functional relations between vehicle currents and 
vehicle densities have attracted much attention, 
which is called the fundamental diagram \cite{Helbing_RMP}.
In our model, we investigate the relation between the molecular  flow rate
and the molecular area fraction under constant external forces.
We define the area fraction $\phi_{}$ as 
the average area fraction in upstream regions ($0<x<150\sigma$),
which corresponds to the accumulated molecular height 
from the middle of the pipe.
Figure \ref{funda} shows the scaled flow rate $q/q_{c}$ as a function of $\phi$
in the pipe with angle $\theta=60^{\circ}$ and the narrowest pipe width 3$\sigma$. 
In each scaled external forces $F/F_{c}$,
scaled flow rates exhibit the linear dependence
of $\phi$ in low area fractions, 
while scaled flow rates decrease in high area fractions.
These behaviors in our system are independent of $F_{c}$ and 
similar to traffic flow behaviors.

The existence of the metastable branch is often discussed in the fundamental diagram of the traffic flow,
while the present model does not exhibit such behaviors.
As the previous subsection has shown, 
we cannot observe any fluctuation peak of the flow in both Figs. \ref{3} and \ref{kakudo_eroo}.
These results likely relate the lack of the metastable branch in our model.
Thus, the flow rate may be represented by a single curve in our fundamental diagram.

\subsection{Width dependence of flow rate}
\begin{figure}
\centerline{
\includegraphics[width=9cm]{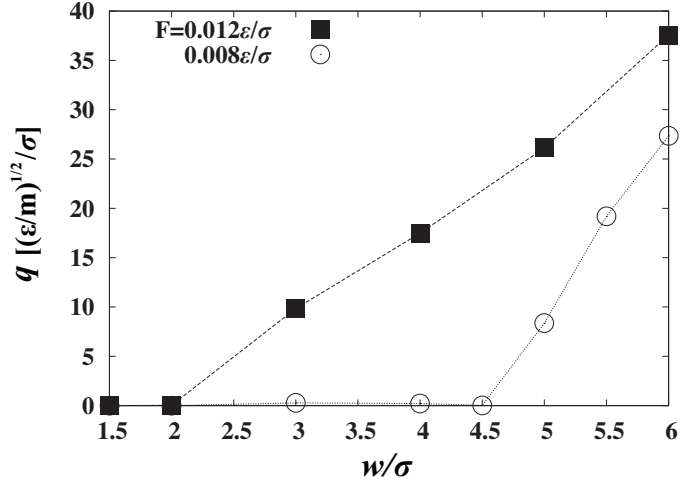}}
\caption{\label{qwidth} Flow rate $q$ as a function of the pipe width at the angle
  $\theta=60^{\circ}$ and  external force $F=0.012\varepsilon/\sigma$ and 
  $0.008\varepsilon/\sigma$.
Dotted lines are connected data.}
\end{figure}
In the situation of gravity-driven granular flow,
jamming transitions depend on the ratio between the exit size and 
the grain size \cite{To}.
Thus, 
we investigate the relation between flow rates and pipe widths 
under constant external forces.
We fix a  pipe configuration with the angle $\theta=60^{\circ}$, but vary the pipe width of the narrowest part 
from $w=1.5\sigma$ to $w=6\sigma$.
The driven particle number is $N=3510$ and the external force is 
$F=0.012\varepsilon/\sigma$, which is above $F_{c}$
and $F=0.008\varepsilon/\sigma$ below $F_{c}$.
Figure \ref{qwidth} shows the flow rate $q$ against the narrowest pipe width $w$.
In $F=0.012\varepsilon/\sigma$, no molecular flow is observed in the pipe width under
 $w=2\sigma$, which 
is too narrow to pass through molecules.
In $F=0.008\varepsilon/\sigma$, we observe no molecular flow below $w=4.5\sigma$
owing to the low external forces.
On the other hand, there exists the flow 
when the width is larger than $w=2\sigma$ for $F=0.012\varepsilon/\sigma$ 
and $w=4.5\sigma$ for $F=0.008\varepsilon/\sigma$.
Although the proportional constant depends on the external force,
we observe that $q$ is almost proportional to the cross
section of the pipe in this flow region.

\subsection{Temperature dependence}
\begin{figure}
\centerline{
\includegraphics[width=9cm]{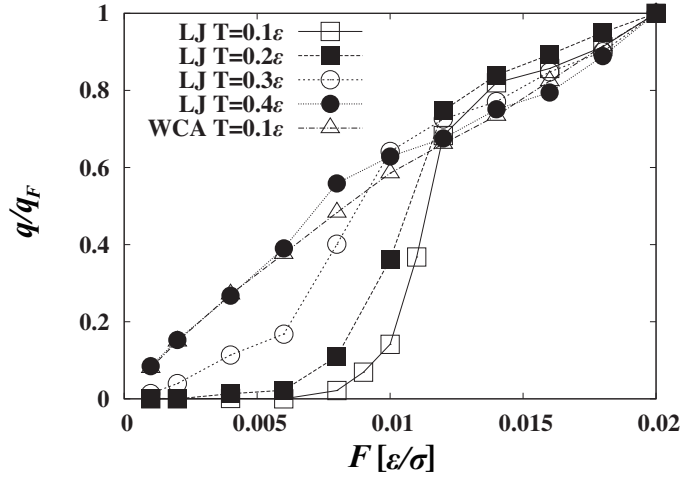}}
\caption{\label{rep}
Scaled flow rate $q/q_{F}$ as a function of the external force in the pipe with angle
$\theta=60^{\circ}$, where $q_{F}$ is the flow rate at $F=0.020 \varepsilon/\sigma$,
for 
LJ molecule at the temperature $T=0.1, 0.2, 0.3, 0.4\varepsilon$
and WCA molecule at the temperature  $T=0.1\varepsilon$.
}
\end{figure}

To investigate the temperature dependence on the flow rate,
we consider LJ molecule at the temperature $T=0.1, 0.2, 0.3, 0.4 \varepsilon$
in the pipe with the angle $\theta=60^{\circ}$ and the width $3\sigma$ and  molecular number $N=3510$.
Figure \ref{rep} shows the scaled flow rate $q/q_{F}$, where $q_{F}$ is the flow rate 
at the external force $F=0.020\varepsilon/\sigma$, as a function of  the external forces.
As we have shown in Fig. \ref{scale}, there is the ``jammed region" for $F<F_{c}$
at the temperature $T=0.1\varepsilon$.
However, as the temperature becomes large, LJ molecules can flow smoothly in the ``jammed" region ($F<F_{c}$)
 and  the flow rate increases linearly against the external force. 
It is known that $T\cong 0.5\varepsilon$ is the transition temperature
from the gas phase to the liquid phase for the wide range of the density \cite{Smit}.
As Fig. \ref{rep} shows, we can observe the smooth  flow at the temperature 
$T \gtrapprox 0.4\varepsilon$,
so that this ``jamming" phenomena likely relate  gas-liquid phase transition.
At the low temperature $T=0.1\varepsilon$,
LJ molecules are regarded as in liquid phase, 
so that the correlated motion of the liquid molecule may cause no molecular flow and dominate the ``jamming" phenomena in the pipe.

\begin{figure}
\centerline{
\includegraphics[width=9cm]{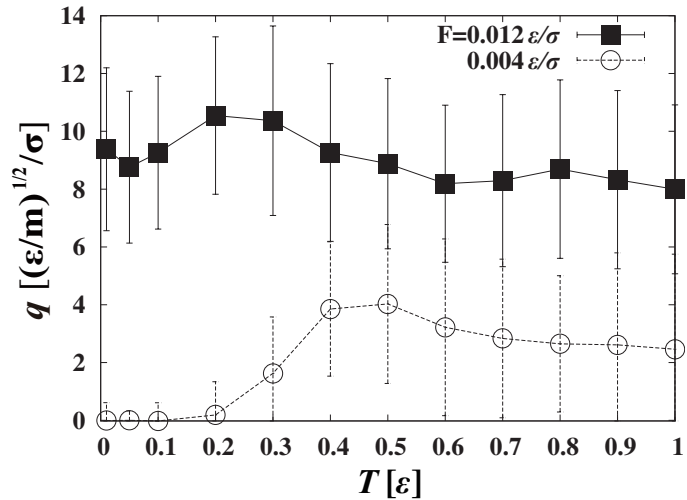}}
\caption{\label{6}
Flow rate $q$ with the standard deviation as a function of the temperature $T$ at the external field $F=0.004\varepsilon/\sigma$ and 0.012$\varepsilon/\sigma$ 
in the pipe with angle $\theta=60^{\circ}$.}
\end{figure}

We also investigate the temperature dependence on the flow in the constant external forces.
Figure \ref{6} shows the flow rate $q$ with the standard deviation  as a function of 
 temperature $T$ in the external field $F=0.004 \varepsilon/\sigma$ 
and 0.012$\varepsilon/\sigma$.
In the case of the external force $F=0.004\varepsilon/\sigma$, 
LJ molecules are almost stuck at the low temperature and begin to flow above about $T=0.2\varepsilon$.
In the case of the large external force $F=0.012\varepsilon/\sigma$, 
LJ molecules flow even in the low temperature
and the ``jamming" phenomena do not happen.
At high temperature, ``jamming" phenomena do not occur  and,  in addition, 
we have observed that "jamming" phenomena depend on the external force.

\subsection{WCA potential}
Thus far, we have investigated the transport of LJ molecules 
that have both attractive and repulsive interactions against other molecules.
To extract the role of attractive interactions, 
we study the molecular flow with WCA potential $\phi_{\text{WCA}}$ in eq. \eqref{WCA}.
As the same condition of LJ molecular flow,
the angle $\theta=60^{\circ}$, the width $3\sigma$,
and  molecular number $N=3510$ are ued.
The plot of the triangle in Fig. \ref{rep}  shows WCA molecule flow.
LJ molecular flow  at the temperature $T=0.1\varepsilon$ is represented by two exponential functions as
mentioned earlier, while
WCA molecular flow can be fitted using a single exponential function.
In the large external forces,
LJ and WCA molecular flows  
exhibit an identical behavior,
since external force is dominant for both flows.

Furthermore, we investigate whether the arch effect appears for the WCA model 
at the pipe with the angle $\theta=90^{\circ}$. 
The WCA molecule can flow smoothly in the pipe in any external force and even 
at  low temperature,
so that  we observe  no arch effect  for the WCA model.
These results indicate that 
the attractive force and low temperature are dominant for "jamming" phenomena.

\section{\label{con}Discussion and conclusions}
We have studied the molecular transport through bottlenecks, 
which is inspired by nanoscale transport phenomena such as biological ion
transport.
In this study, we simplify the bottleneck as the two-dimensional pinched pipe
characterized by the angle $\theta$ and width $w$.
As the external forces increase, we can observe two distinct molecular flow
regimes.
Using the value at the crossover of these regimes,
which is considered to correspond to the balance
of the molecular interaction and the external force, 
the scaling behaviors of the flow rate and external force 
are observed in our model.
These scaling behaviors show that the flow rate  
does not depend on the pipe configurations at  the angle below $\theta=80^{\circ}$. 
However, at the large angle, we cannot clarify the relation between the flow rate
and external force
owing to the small amount of  molecular flow.
The flow rate as a function of the angle is also observed, and this relation
is verified using  a simple theory.
We further exhibit the value of the critical flow rate from rough theoretical estimations.
These double exponential forms of the flow rate against the external force
can be clearly observed at the low temperature $T=0.1\varepsilon$.
On the other hand, as the temperature increases,
``jamming" region disappears even in the small external force
and this ``jamming" phenomena are likely related to 
the gas-liquid transition of the 2D LJ phase diagram.
In the fundamental diagram,  
the change in the fluctuation in the metastable region is often discussed,
whereas we cannot observe any fluctuation peak in our model.
These results likely relate to the lack of the metastable branch and thus
the flow rate may be represented by a single curve in our fundamental diagram.
We also investigated the WCA molecular flow  to extract the role of the
attractive forces and found that 
attractive forces at the low temperature  are dominant for the ``jamming'' phenomena.

The extension of this model to other transport models is straightforward.
As examples, 
grand canonical molecular simulations have been developed to 
examine the molecular transport by introducing explicit chemical potential
gradients into the system \cite{cracknell,Heffelfinger}.

As emphasized in the introduction, we study a very simple model of
the pipe with a  bottleneck,
while our model qualitatively shows the important features of the molecular
transport
through the bottleneck and is of importance for the architecture and 
functionality in the emerging field of nanomaterial systems.

\section*{Acknowledgements}
The authors thank Hirofumi Wada for fruitful suggestions 
and Glenn Paquette for correcting proofs.
The numerical calculations were carried out on Altix3700 BX2 at YITP of Kyoto
University and this work was supported by the Grant-in-Aid for the Global COE Program "The Next Generation of Physics, Spun from Universality and Emergence" from the Ministry of Education, Culture, Sports, Science and Technology (MEXT) of Japan.
This work is also supported by Grants-in-Aid of MEXT(Grant Nos. 21015016 and 21540384)

%

\end{document}